\documentclass[%
 reprint,
 amsmath,amssymb,
 aps,nofootinbib,   
]{revtex4-1}
\usepackage{bm}
\PassOptionsToPackage{linktocpage}{hyperref}
\usepackage[hyperindex,breaklinks]{hyperref}
\usepackage{enumitem}
\usepackage{slashed}

\renewcommand{\theta}{\vartheta}

\usepackage{array}
\usepackage{mathtools}

\usepackage{etoolbox}

\begin{document} 

\title{A Proof of the Axion?}

\author{Gia Dvali$^{a,b,c}$,  Cesar Gomez$^{d}$  and Sebastian Zell$^{a,b}$} 
\affiliation{
	$^a$Arnold Sommerfeld Center, Ludwig-Maximilians-Universit\"at, Theresienstra{\ss}e 37, 80333 M\"unchen, Germany
}
\affiliation{
	$^b$Max-Planck-Institut f\"ur Physik, F\"ohringer Ring 6, 80805 M\"unchen, Germany
}
\affiliation{ 
	$^c$Center for Cosmology and Particle Physics, Department of Physics, New York University, 726 Broadway, New York, NY 10003, USA
}
\affiliation{$^d$Instituto de F\'{\i}sica Te\'orica UAM-CSIC, Universidad Aut\'onoma de Madrid, Cantoblanco, 28049 Madrid, Spain}


\begin{abstract}
We show that the {\it de Sitter quantum breaking bound } when applied to 
QCD exposes the necessity of the axion solution to the strong CP 
problem. 
The Peccei-Quinn  mechanism emerges as a {\it consistency} requirement 
independent of the naturalness questions. The $\theta$-angle must be unphysical rather than simply small. 
 All other approaches including a fine-tuning of $\theta$
lead to the existence of de Sitter vacua and are excluded by consistency.

 \end{abstract}


\maketitle

 It has become evident that a theory with either local or global  potential minima with positive energy density is exposed to a   
fundamental problem of {\it de Sitter quantum breaking}  \cite{us1}.  
In the former work this led us to the formulation of a criterion  requiring that such vacua must be absent in any consistent theory. Below we shall adopt this requirement.

  Our bound fully resonates with the recently proposed de Sitter 
 swampland conjecture  
 \cite{swamp1} (see \cite{us2} for the connection 
between the two proposals and references therein 
for other discussions).  

 The basic consistency requirement of \cite{us1} is that a positive vacuum energy in any de Sitter like Hubble patch must be relaxed by a (scalar)  degree of freedom during a time $t_{esc}$ that needs to be shorter than the quantum break-time $t_Q$.  
  The latter depends 
 on the interaction strength. An absolute upper bound 
 due to gravity is $t_Q = M_P^2 / (N_s H^3)$, where 
 $M_P$ is the Planck mass, $H$ is the Hubble parameter of the would-be de Sitter state and $N_s$ is the number of active particle species. 
 
  As already discussed in \cite{us1}, an immediate consequence of
  the de Sitter quantum breaking bound is that the currently observed  dark energy cannot be constant and must change in time.  We shall not specify 
 the precise form of its evolution but only assume that it drives our Universe  towards Minkowski vacuum within a time shorter than $t_Q$.  This is an extremely mild restriction since for the current phase the quantum
 break-time from gravity is enormous, $t_Q \sim 10^{134}\ \text{y}$.\\   
    
     In an accompanying article \cite{us5} we show that the quantum breaking bound severely constrains some extensions of the Standard Model. 
      In the present note we shall continue this line of research and demonstrate that the same bound when applied to QCD essentially proves the necessity of 
  rendering the $\theta$-angle unphysical by the axion, {\it regardless} 
  of any naturalness  concerns.  That is, the Peccei-Quinn mechanism \cite{PQ}
 must be at work as a {\it consistency requirement}.  No other approach, including the {\it fine tuning} of $\theta$, is acceptable.

  The argument is very simple and goes as follows.  Consider 
  QCD with no axions present.  In this case 
  the theory contains vacua belonging to different superselection sectors 
  that are labeled by the $\theta$-parameter \cite{vac}. Let the energy dependence 
  on it be $E(\theta)$.  
  Now, we put together the following three facts (two experimental and one theoretical). 
 First,  the experimental
 upper bound on $\theta$ is minuscule, below $10^{-10}$ or so. 
  Secondly, we live in a state with a very small and positive dark energy 
density $\sim (10^{-3}\ \text{eV})^4$ (as said above, according to our criteria this itself cannot be a constant).   
  Thirdly, we know from the Vafa-Witten theorem \cite{VW} that 
  the global minimum of the energy is achieved at $\theta=0$. 
  
  Putting together the above three facts immediately implies that if $\theta$ is physical, our $\theta$-vacuum 
is very close to the one with the lowest energy.   
 Then, there must exist plenty of vacua 
with positive energy densities larger than ours. These are the 
$\theta$-vacua corresponding to the 
interval $10^{-10} \lesssim \theta \lesssim \pi$. 

 Ignoring a mild suppression by the light quark masses,   
the highest among the $\theta$-vacua would have the energy density set by the QCD scale
  $\Lambda_{QCD}^4 \sim (\text{GeV})^4$.  
   All these vacua are of de Sitter-type. Hence they are 
  exposed to quantum breaking \cite{us1} and therefore are inconsistent.  Thus, such a possibility is excluded by our criterion. This leaves us with the only option 
  that $\theta$ must be unphysical due to the existence of an axion.\\
   
  Once the axion is introduced, no problem with quantum breaking persists. 
  The points $\theta \neq 0$ are no longer vacua and swiftly relax into the 
  state $\theta =0$ before any quantum breaking can set in. 
  Indeed, the relaxation time is set by the inverse axion mass,
  $t_{esc} \sim m_a^{-1}  \sim f_a / \Lambda_{QCD}^2$, 
  with $f_a$ being the axion decay constant. This  relaxation time is much shorter than the 
  corresponding  quantum break-time of any would-be de Sitter
  state with $\theta \neq 0$.  The highest energy state would have the shortest 
  break-time given by  $t_Q \sim M_P^5 / \Lambda_{QCD}^6$, which can be represented as  $t_Q \sim t_{esc}\, M_P^5 / (\Lambda_{QCD}^4 f_a) $.  Even for $f_a \sim M_P$  this gives 
  $t_Q \sim 10^{76}\, t_{esc}$ and thus no quantum breaking happens.  
  After relaxation the quantum field continues to perform damped oscillations 
 without encountering any inconsistency with quantum breaking \cite{axionQB}. \\
  
  It is remarkable that an intrinsic quantum inconsistency of de Sitter
 imposes the necessity of the axion once we couple QCD to gravity.  This fact singles out the Peccei-Quinn mechanism as 
 a consistency requirement irrespective of whether one cares about the naturalness of small $\theta$. In other words, it is not sufficient for 
 $\theta$ to be small, it must be literally {\it unphysical}. \\

{\bf Acknowledgements.}
This work was supported in part by the Humboldt Foundation under Humboldt Professorship Award and ERC Advanced Grant 339169 "Selfcompletion".

\end{document}